\documentclass{article}

\usepackage{amsmath}
\usepackage{PRIMEarxiv}
\usepackage{mathrsfs}
\usepackage{listings}
\usepackage{longtable}
\usepackage{tabularx}
\usepackage{rotating}
\usepackage{algorithm}
\usepackage{algorithmic} 
\usepackage{adjustbox}
\usepackage{forloop}
\usepackage{makecell}
\usepackage{etoolbox}
\usepackage{multirow}
\usepackage{subcaption}
\usepackage{float}
\usepackage{calc}
\usepackage{wrapfig,lipsum,booktabs} 
\usepackage{xtab} 
\usepackage[dvipsnames]{xcolor}
\usepackage{tikz}
\usetikzlibrary{arrows.meta, positioning, shapes.geometric}

\usepackage[utf8]{inputenc} 
\usepackage[T1]{fontenc}    
\usepackage{hyperref}       
\usepackage{url}            
\usepackage{booktabs}       
\usepackage{amsfonts}       
\usepackage{nicefrac}       
\usepackage{microtype}      
\usepackage{lipsum}
\usepackage{fancyhdr}       
\usepackage{graphicx}       
\graphicspath{{media/}}     

\title{Sentiment-Aware Stock Price Prediction with Transformer and LLM-Generated Formulaic Alpha}

\author{
  Qizhao Chen \\
  Graduate School of Information Science \\
  University of Hyogo \\
  Kobe, Japan\\
  \texttt{af24o008@guh.u-hyogo.ac.jp} \\
  \AND
  Hiroaki Kawashima \\
  Graduate School of Information Science \\
  University of Hyogo \\
  Kobe, Japan\\
  \texttt{kawashima@gsis.u-hyogo.ac.jp} \\
}

\begin{document}
\maketitle

\begin{abstract}
Traditionally, traders and quantitative analysts address alpha decay by manually crafting formulaic alphas, mathematical expressions that identify patterns or signals in financial data, through domain expertise and trial-and-error. This process is often time-consuming and difficult to scale. With recent advances in large language models (LLMs), it is now possible to automate the generation of such alphas by leveraging the reasoning capabilities of LLMs. This paper introduces a novel framework that integrates a prompt-based LLM with a Transformer model for stock price prediction. The LLM first generates diverse and adaptive alphas using structured inputs such as historical stock features (Close, Open, High, Low, Volume), technical indicators, sentiment scores of both target and related companies. These alphas, instead of being used directly for trading, are treated as high-level features that capture complex dependencies within the financial data. To evaluate the effectiveness of these LLM-generated formulaic alphas, the alpha features are then fed into prediction models such as Transformer, LSTM, TCN, SVR, and Random Forest to forecast future stock prices.  Experimental results demonstrate that the LLM-generated alphas significantly improve predictive accuracy. Moreover, the accompanying natural language reasoning provided by the LLM enhances the interpretability and transparency of the predictions, supporting more informed financial decision-making.
\end{abstract}


\section{Introduction}\label{sec1}

Financial markets are complex systems where prices are influenced by company fundamentals, technical patterns, sentiment, and external events. Predicting stock prices remains a major challenge in both finance and machine learning. Traditional models like regression and tree-based methods often struggle to capture the highly non-linear relationships present in financial time series.

Deep learning models, especially Transformers, have recently shown strong results in sequence modeling tasks. Their ability to learn long-range dependencies makes them suitable for predicting price movements~\cite{9731073,electronics13214225,10185006,Chen2025LLMTransformer}. However, using only raw financial features often leads to noisy and unstable predictions. To improve model performance, many researchers and practitioners create special features called "alpha factors" or "alphas." These are formulas designed to capture patterns that can predict future returns~\cite{chen2025adaptivealphaweightingppo}. Equation~\ref{eq: alpha exmaple} is an example of a formulaic alpha.

\begin{equation}
\label{eq: alpha exmaple}
\text{Alpha}_t = \left( \frac{P_t - \text{SMA}_{20,t}}{\text{SMA}_{20,t}} \right) \cdot \log(1 + \text{Sentiment}_t)
\end{equation}

This alpha factor combines a technical indicator with sentiment information. The term
\( \frac{P_t - \text{SMA}_{20,t}}{\text{SMA}_{20,t}} \) captures the percentage deviation of the current price \( P_t \) from its simple moving average of 20 days, indicating short-term overbought or oversold conditions. The logarithmic sentiment term \( \log(1 + \text{Sentiment}_t) \) adjusts this signal based on the sentiment of the market for the day, where higher sentiment values amplify the signal and negative or neutral sentiment dampens it. By integrating both technical momentum and investor mood, this alpha aims to forecast short-term returns more effectively than using technical or sentiment data alone.

In practice, alpha factors do not remain effective forever. Over time, as more traders discover and exploit an alpha, it tends to lose its predictive power. This phenomenon is known as alpha decay~\cite{doi:10.1287/mnsc.2022.4353}. Alpha decay is a major problem in quantitative finance because it forces investors to constantly search for new and original alphas. Relying on some old alphas can quickly lead to underperformance as markets evolve and adapt.

Traditionally, designing new alphas has been a manual process, driven by human intuition and experience. Analysts and quants study historical patterns, test hypotheses, and refine formulas through trial and error. However, this approach is slow, expensive, and limited by human creativity. In recent years, Large Language Models (LLMs) like GPT have shown strong abilities in reasoning, summarization, and idea generation. This opens up the possibility of using LLMs to automate the search for new alphas by generating formulaic trading signals based on structured financial data~\cite{tang2025alphaagentllmdrivenalphamining}.

Prompt-based LLMs offer several advantages in this context. First, they can quickly generate a large number of candidate formulas following clear instructions (prompts). Second, the generated alphas are often interpretable because they are written in a human-readable form, unlike the hidden features learned by deep neural networks. Third, prompt-based LLMs are flexible. By adjusting the prompt design, users can guide the LLM to focus on different aspects of the market, such as momentum, mean reversion, or sentiment. This flexibility allows for fast exploration of many trading ideas without heavy manual effort. Lastly, since LLMs can be fine-tuned or adapted over time, they can continuously assist in generating fresh alphas to fight alpha decay.

In this paper, we propose a new framework that uses prompt-based LLMs to automatically generate formulaic alphas from stock features, technical indicators, and sentiment scores. These generated alphas serve as additional inputs to a vanilla Transformer model (encoder-decoder structure)~\cite{NIPS2017_3f5ee243} that predicts next-day stock prices. By introducing new machine-generated alphas, it is expected that alpha decay will slow down and stronger predictive signals will be maintained over time.

Following our previous work~\cite{chen2025sentiment}, we also include the sentiments of related companies as one of the input features. Related companies are defined as those that have connections with the target companies, such as business partners, suppliers, or competitors. The reason is that the news sentiments of these companies can also affect the stock movement of the target company. In our earlier study, we applied Named Entity Recognition (NER) to extract company names from news articles about target firms. The underlying assumption is that companies frequently co-occurring in the same articles as the target company are likely to have some form of business relationship or relevance.

In summary, this paper addresses the following two research questions (RQs):

\begin{itemize}

\item RQ1: Can sentiment information from related companies be effectively incorporated into alpha formula generated by LLM?
\item RQ2: Does sentiment-aware LLM-generated alpha signals improve stock price prediction accuracy using deep learning techniques such as Transformer?

\end{itemize}

The remainder of this paper is structured as follows. Firstly, related work is listed (Section 2). The methodology is then described (Section 3). In Section 4, the experimental results are presented. Section 5 presents further discussion of the findings. Finally, Section 6 concludes the paper.

\section{Related Work}\label{sec2}

\subsection{Stock Price Prediction with Machine Learning Models}

Predicting stock prices has attracted a lot of research attention. Traditional methods include statistical models such as ARIMA~\cite{7046047} and GARCH~\cite{MUTINDA2024e02374}. These models assume certain patterns in the data, such as stationarity or constant volatility, but real financial data often violates these assumptions. As a result, these models struggle when applied to complex market environments.

With the rise of machine learning, models such as Linear Regression~\cite{10381982}, Support Vector Machines~\cite{6703096}, Random Forests~\cite{9987903}, and Gradient Boosting~\cite{10915343} have been widely used for stock prediction. These models can capture non-linear patterns better than traditional statistical approaches. However, they still require careful feature engineering, and their ability to model sequential data is limited.

Deep learning methods have shown strong promise for stock prediction. Recurrent Neural Networks (RNNs) and Long Short-Term Memory (LSTM) networks were among the first deep learning models used for financial time series~\cite{8950831,10392023,MOGHAR20201168}. They are designed to capture temporal dependencies in data. While LSTMs improve over RNNs by addressing the vanishing gradient problem, they are still limited when handling long sequences.

Transformers, introduced later, overcame these limitations by relying on self-attention mechanisms. Transformers can capture long-term dependencies more effectively and have become a popular choice for time series forecasting tasks, including in finance. Several studies have adapted the Transformer architecture for stock prediction by treating historical prices, technical indicators, and other features as sequential inputs~\cite{10825946,li2025transformerbasedtimeseriesforecasting,Chen2025AnomalyAware,chen2025comparingdifferenttransformermodel}. Some approaches modify the attention mechanism to better focus on important market events~\cite{kaeley2023supportstocktrendprediction}. Others combine Transformers with other models, such as Convolutional Neural Networks (CNNs), to capture both short-term and long-term patterns~\cite{zeng2023financialtimeseriesforecasting,CNN-Trans-SPP}. The flexibility of Transformers in handling different types of input data makes them attractive for financial forecasting.

\subsection{Feature Engineering and Formulaic Alphas}

Feature engineering remains a key part of the building of successful stock prediction models. Many researchers have proposed the construction of technical indicators such as moving averages, RSI, MACD, and Bollinger bands as features~\cite{9580172,8806422,deep2025riskadjustedperformancerandomforest}. Some works also use fundamental data~\cite{phan2024leveragingfundamentalanalysisstock,9688449}, such as earnings reports and balance sheet ratios, while others explore alternative data sources such as social media posts and news sentiment~\cite{10112768}.

The idea of alpha factors is central to quantitative finance. Early works manually crafted alphas based on theories such as momentum, mean reversion, and value investing~\cite{FAMA19933}. Later, research moved toward automating alpha discovery by using genetic algorithms and symbolic regression to search for profitable formulas~\cite{ren2024alphaminingenhancingwarm}. More recently, some studies proposed the use of reinforcement learning to generate or select alphas, in order to reduce human bias and improve performance~\cite{zhao2024quantfactorreinforceminingsteady,Yu_2023}. Despite these efforts, alpha decay remains a persistent issue, as once powerful alphas often weaken when exposed to public markets.

Large Language Models (LLMs) have opened new possibilities for finance. LLMs have been used to summarize earnings calls, predict market sentiment from news articles, and even generate trading strategies~\cite{Chen2025}. However, using LLMs to generate formulaic alphas is still a new area. Prompt-based approaches allow users to control the output of LLMs through designed instructions. This method can guide the LLM to create structured and interpretable trading formulas instead of free-form text. It combines human creativity and machine efficiency, making it attractive for alpha generation.

Unlike previous studies such as~\cite{tang2025alphaagentllmdrivenalphamining}, which use alphas to generate direct trading signals, this study treats alphas as high-level structured features that capture complex relationships among financial inputs. These structured features are then used as inputs to prediction models for forecasting stock prices.

\section{Methodology}\label{sec3}

This section explains the data sources, preprocessing steps, model design, and training methods used in this study. We aim to combine traditional financial features, sentiment analysis, and prompt-based LLM-generated alphas to predict stock prices. 

\begin{figure*}[!htbp]
    \centering
    \includegraphics[width=1\textwidth]{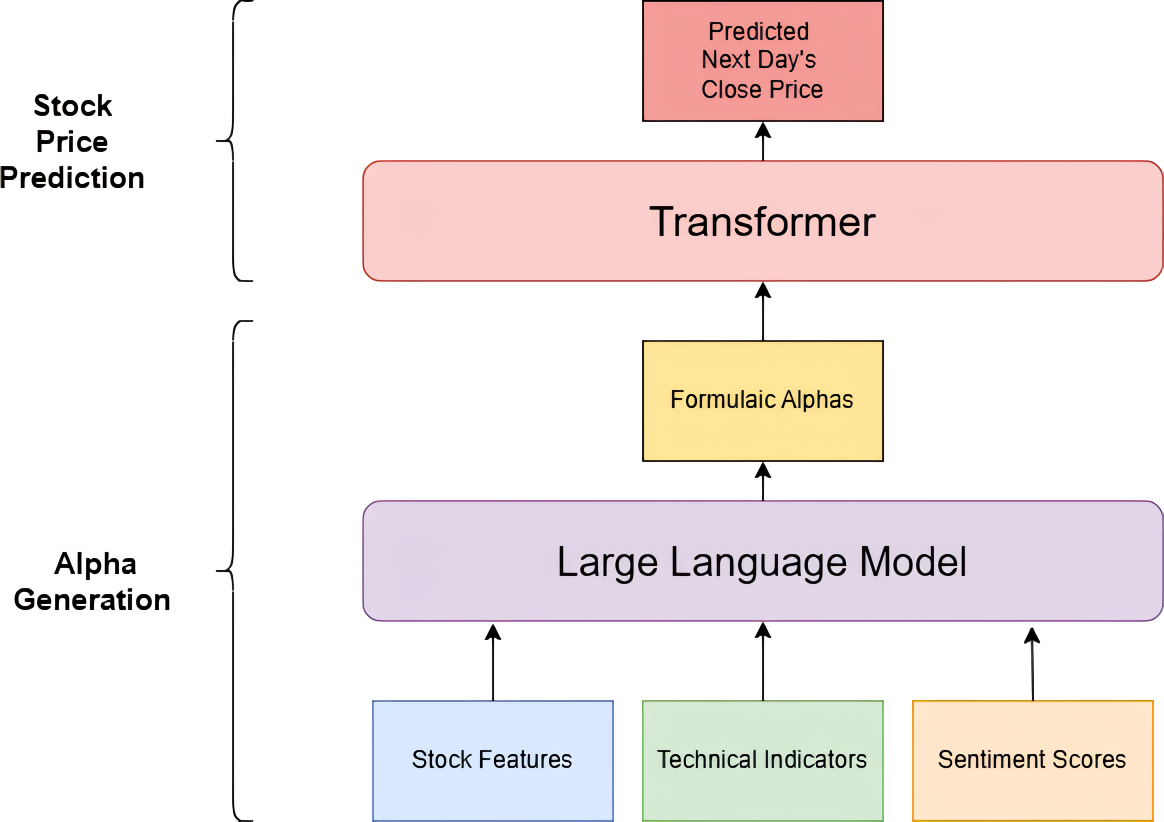}  
    \caption{Whole Picture of the Proposed Method}
    \label{fig: whole picture}
\end{figure*}

Figure~\ref{fig: whole picture} illustrates the framework proposed in this study. First, the LLM uses basic features such as stock features and technical indicators, together with sentiment scores of related companies, to generate formulaic alphas, which then feed a Transformer model to predict the stock price in the next day.

\subsection{Data}
Following common practice in related stock prediction studies, which often evaluate models using a limited set of representative stocks (typically between one to ten)~\cite{computation13010003, VIJH2020599,ABOLMAKAREM2024200449,electronics13173396,CHEN2023106038,10.1007/978-981-16-7305-4_15}, this study selects five companies for empirical analysis. This choice allows for a focused evaluation of the proposed method while maintaining manageable computational complexity. The stock data for five companies, Apple, HSBC, Pepsi, Toyota, and Tencent, is collected using the yfinance Python library. The data cover the period from May 1, 2016, to May 8, 2024. These companies are chosen because they operate in different industries and markets, including technology, finance, consumer goods, automotive, and internet services, which allows the proposed framework to be evaluated under diverse market characteristics. All five stocks are highly liquid and widely followed, ensuring data availability. The stock data includes Open, High, Low, Close, and Volume information.

Technical indicators are calculated using the pandas-ta library. Technical indicators are mathematical calculations based on stock price, volume, or open interest. They are widely used by traders to identify market trends and predict future price movements. Table~\ref{tab:technical_indicators} shows the technical indicators used in this study. The \textbf{Simple Moving Average (SMA)} smooths out short-term price movements by taking the average closing price over a fixed number of days. It helps identify basic trends over time. The \textbf{Exponential Moving Average (EMA)} gives more weight to recent prices, making it respond faster to new market changes than the SMA. \textbf{Momentum} shows how quickly the price is changing by comparing the current price to a price from a few days ago. It signals the strength of a price move. The \textbf{Relative Strength Index (RSI)} measures how fast and strong price changes are, helping spot if a stock is overbought or oversold. The \textbf{Moving Average Convergence Divergence (MACD)} tracks the difference between two EMAs. It highlights shifts in trend strength and direction. \textbf{Bollinger Bands} set upper and lower limits around a moving average. When prices move outside the bands, it may suggest high volatility or a reversal. \textbf{On-Balance Volume (OBV)} adds or subtracts trading volume based on price direction. It shows how volume supports or contradicts price moves.

\begin{table}[h]
\centering
\caption{List of Technical Indicators Used}
\begin{tabular}{ll}
\hline
\textbf{Indicator} & \textbf{Description} \\
\hline
SMA\_5 & Simple Moving Average (5 periods) \\
SMA\_20 & Simple Moving Average (20 periods) \\
EMA\_10 & Exponential Moving Average (10 periods) \\
Momentum\_3 & Momentum Indicator (3 periods) \\
Momentum\_10 & Momentum Indicator (10 periods) \\
RSI\_14 & Relative Strength Index (14 periods) \\
MACD & MACD Line (fast=12, slow=26) \\
MACD\_Signal & MACD Signal Line (signal=9) \\
BB\_Upper & Bollinger Bands Upper (20-period, 2 std) \\
BB\_Lower & Bollinger Bands Lower (20-period, 2 std) \\
OBV & On-Balance Volume \\
\hline
\end{tabular}
\label{tab:technical_indicators}
\end{table}

In addition to stock data, daily news articles related to the five companies as well as their related companies are downloaded from websites such as Yahoo News using the Financial News Feed and Stock News Sentiment Data API\footnote{https://eodhd.com/financial-apis/stock-market-financial-news-api}. All news is in English. Based on our previous study, related companies are defined as those that appear frequently in news articles about the five target companies~\cite{chen2025sentiment}. For example, BYD appears in Toyota’s news articles over one hundred times, suggesting a connection between the two companies. The news articles help capture market sentiment, which can have an important impact on stock prices. Previously, we used NER to extract entity names from news articles related to the target companies, based on their frequency of appearance. All identified listed companies were then ranked from highest to lowest frequency, and for each target company, the top ten most frequently co-mentioned companies were selected as related companies. The Python library spaCy is used to perform NER. Figure~\ref{fig:ner_cooccurrence} demonstrates the whole process of the related company selection. Table~\ref{tab: Related Firms} shows the target companies and their corresponding related companies. Furthermore, the related companies were extracted using the whole news dataset. However, the extraction step is performed once and produces a fixed set of related companies that is independent of the model training and prediction process. During training and testing, the model only uses news and price information from the corresponding time periods, and no test-period news content is used to update model parameters or input features. As a result, future information is not exploited in the predictive task. Figure~\ref{fig:no_leakage} shows the related company extraction and model training procedure.

\begin{figure}[htbp]
\centering
\begin{adjustbox}{max width=\linewidth}
\begin{tikzpicture}[
    node distance=1.8cm and 2.5cm,
    every node/.style={
        draw, rectangle, rounded corners,
        align=center,
        minimum height=1cm,
        minimum width=2.8cm
    },
    arrow/.style={->, thick}
]

\node (news) {News Articles};
\node (ner) [right=of news] {Named Entity\\Recognition (NER)};
\node (entities) [right=of ner] {Extracted Company\\Names};

\node (cooccur) [below=of entities] {Co-occurrence\\Frequency Count};
\node (select) [right=of cooccur] {Related Company\\Selection};

\draw[arrow] (news) -- (ner);
\draw[arrow] (ner) -- (entities);
\draw[arrow] (entities) -- (cooccur);
\draw[arrow] (cooccur) -- (select);

\node[draw=none, font=\small, below=0.3cm of ner]
{Identify organization entities};

\node[draw=none, font=\small, below=0.3cm of cooccur]
{Count joint appearances within the news article};

\end{tikzpicture}
\end{adjustbox}
\caption{Company extraction using named entity recognition and related company selection based on co-occurrence frequency.~\cite{chen2025sentiment}}
\label{fig:ner_cooccurrence}
\end{figure}

\begin{figure}[htbp]
\centering
\begin{adjustbox}{max width=\textwidth}
\begin{tikzpicture}[
    box/.style={rectangle, draw, rounded corners, minimum width=3.5cm, minimum height=1cm, align=center},
    arrow/.style={->, thick},
    dashedbox/.style={rectangle, draw, dashed, rounded corners, minimum width=4.2cm, minimum height=1.2cm, align=center}
]

\node[box] (news) {Daily News Dataset};
\node[box, below=1.4cm of news] (extract) {Related-Company Extraction \\ (NER)};
\node[box, below=1.4cm of extract] (company) {Fixed Related-Company Lists \\ Example List: \\ Apple \\ Amazon \\ Meta \\ Google};

\node[dashedbox, right=4.8cm of extract] (split) {Temporal Split\\Training / Testing};

\node[box, below=1.4cm of split] (train) {Model Training\\(with training period only) \\ Example: \\ Apple\_polarity: 0.15 \\ Amazon\_polarity: 0.56 \\ Meta\_polarity: -0.8 \\ Google\_polarity: 0 
\\(Google is absent from the news)};
\node[box, below=1.4cm of train] (test) {Model Evaluation\\(with testing period only) \\ Example: \\ Apple\_polarity: 0.25 \\ Amazon\_polarity: 0.06 \\ Meta\_polarity: -0.28 \\ Google\_polarity: 0.17
\\(Google appears in the news)};

\draw[arrow] (news) -- (extract);
\draw[arrow] (extract) -- (company);

\draw[arrow] (news.east) -- ++(1.2cm,0) |- (split.west);
\draw[arrow] (split) -- (train);
\draw[arrow] (train) -- (test);

\node[align=center, text width=4cm, below=0.8cm of company] {\small Related-company lists are\\ extracted once and fixed};

\end{tikzpicture}
\end{adjustbox}
\caption{Illustration of related-company extraction and model training procedure without data leakage. The related-company list is extracted once and kept fixed. During model training, if a company does not appear in the training-period news
(e.g., Google), its polarity value is set to zero, which will not cause information leakage during training.}
\label{fig:no_leakage}
\end{figure}

In this study, the extracted related companies will be further analyzed by the LLM to identify the most influential ones that have a strong connection with stock movement. The company selected by the LLM will then be incorporated into the formulaic alphas.

\begin{algorithm}[H]
\caption{Time-Series Stock Prediction Pipeline}
\label{alg:stock_pipeline} 
\begin{algorithmic}
\STATE \textbf{Input:} Daily news and stock dataset $D$, sliding window size $w$, prediction horizon $h$
\STATE \textbf{Output:} Trained model and test predictions

\STATE Split $D$ chronologically into training set $D_{train}$ and testing set $D_{test}$

\STATE Extract related-company lists from $D$ (fixed, no update during training/testing)

\STATE Compute LLM-generated features for all companies

\STATE Fit MinMax scalers on $D_{train}$ features and labels
\STATE Apply scalers to $D_{train}$ and $D_{test}$ (transform only)

\STATE Construct sliding windows of length $w$:
\FOR{each time $t$ in $D_{train}$}
    \STATE $X_{train}[t] \gets$ features from $t-w$ to $t-1$
    \STATE $y_{train}[t] \gets$ close price at $t+h$
\ENDFOR

\FOR{each time $t$ in $D_{test}$}
    \STATE $X_{test}[t] \gets$ features from $t-w$ to $t-1$
    \STATE $y_{test}[t] \gets$ close price at $t+h$
\ENDFOR

\STATE Train prediction model (e.g., Transformer, TCN) on $(X_{train}, y_{train})$

\STATE Evaluate model on $(X_{test}, y_{test})$

\RETURN Trained model and test predictions
\end{algorithmic}
\end{algorithm}

\begin{table*}[!htbp]
    \centering
    \caption{Related Companies~\cite{chen2025sentiment}}
    \label{tab: Related Firms}
    \renewcommand{\arraystretch}{2}
    \begin{tabular}{|c|p{10cm}|}
    \hline
    \textbf{Target Company} & \textbf{Related Company} \\
    \hline
    Apple & Google, Amazon, Microsoft, Tesla, Samsung, Nvidia, Meta, Intel, AMD, IBM \\
    \hline
    HSBC & JPMorgan, Standard Chartered, Morgan Stanley, UBS, Citigroup, Goldman Sachs, Credit Suisse, NatWest, Apple, Microsoft \\
    \hline
    Tencent & NetEase, Baidu, JD, Alibaba, Microsoft, Apple, Google, Amazon, Meta, Nvidia, Intel \\
    \hline
    Toyota & Nissan, Honda, Mazda, Ford, GM, Volkswagen, BMW, Tesla, Hyundai, BYD, Apple, Amazon \\
    \hline
    Pepsi & Coca-Cola, Walmart, Costco, Colgate, Chevron, P\&G, Johnson \& Johnson, Apple, Microsoft, Amazon, Tesla, IBM \\
    \hline
    \end{tabular}
\end{table*}

\begin{table*}[h]
\centering
\caption{Structured Prompt for Alpha Generation}
\label{tab:alpha_prompt}
\renewcommand{\arraystretch}{1.5} 
\begin{tabular}{|p{\textwidth}|}
\hline
\textbf{Task Prompt: Generating Predictive Alphas for \{\textcolor{blue}{Company}\}’s Stock Prices} \\
\hline
\textbf{Objective:} Generate formulaic alpha signals to predict \{\textcolor{blue}{Company}\}’s stock prices using: \\
1. \textbf{Stock features} (e.g., close, open, high, low, volume), \\
2. \textbf{Technical indicators} (e.g., RSI, moving averages, MACD), \\
3. \textbf{Sentiment data} for target company and its related companies (e.g., competitors, business partners, suppliers). \\

\textbf{Input Data:} \\
A single \texttt{pandas.DataFrame} \{\textcolor{blue}{DataFrame Input in JSON Format}\} with rows representing trading days and columns including: \\
- \textbf{Stock Features:} \texttt{Close}, \texttt{Open}, \texttt{High}, \texttt{Low}, \texttt{Volume} \\
- \textbf{Technical Indicators:} e.g., \texttt{RSI}, \texttt{SMA}, \texttt{EMA}, \texttt{MACD}, \texttt{Bollinger Bands} \\
- \textbf{Sentiment Data:} Daily sentiment polarity scores for the target company and related companies.
\\

\textbf{Requirements:} \\
1. \textbf{Alpha Formulation:} Propose 5 formulaic alphas combining stock features, cross-company sentiment divergences or technical indicators. \\
2. \textbf{Feature Engineering:} Normalize inputs (e.g., Z-scores), handle missing data. \\

\textbf{Example Alpha:} \\
$\alpha_1 = \text{Apple\_5D\_Return} + 0.5 \times (\text{Apple\_Sentiment} - \text{Google\_Sentiment})$ \\
\hline
\end{tabular}
\end{table*}

Each news article is processed to calculate a polarity score. The polarity score is calculated using VADER (Valence Aware Dictionary and sEntiment Reasoner)~\cite{Hutto_Gilbert_2014}, which is a lexicon- and rule-based sentiment analysis tool specifically designed to analyze sentiments expressed on social media and short texts. It works by using a predefined dictionary (lexicon) that maps words to their sentiment intensity scores, positive, negative, or neutral. The polarity score measures how positive, neutral, or negative a piece of text is. A positive score indicates optimistic sentiment, a negative score reflects pessimism, and a score near zero suggests neutral sentiment. Daily polarity scores are averaged to obtain a single sentiment score for each company per day. These sentiment scores are then used as part of the input features.

All timestamps for the stock and news data are recorded in UTC to avoid the time zone inconsistency between different stocks. The news data used in this study are at the daily level, and intraday timestamps are not used. Each day’s polarity score is computed solely from news published on that calendar day.

Following the temporal order of the time series, the first 70\% of the data is used for training and the remaining 30\% of the data is used for testing. A sliding window of 5 days is applied to the input features. The stock prediction pipeline is illustrated in Algorithm~\ref{alg:stock_pipeline}.

\subsection{Formulaic Alpha Generation Using LLM}
To generate new alpha formulas, we use the deepseek-r1-distill-llama-70b model, which is a distilled variant of the DeepSeek-R1 model based on the LLaMA architecture and is provided by Groq\footnote{\url{https://groq.com/}} for inference and deployment. Knowledge distillation is a technique in which a smaller model (student) learns to mimic the behavior of a larger, more powerful model (teacher). This process reduces the size and computational cost of the model while keeping much of the original performance.

The LLM is controlled through prompting. In this study, the prompt (Table~\ref{tab:alpha_prompt}) guides the LLM to generate formulaic alphas. Specifically, the prompt is a structured instructional task prompt designed to instruct the model to generate formulaic alphas for stock price prediction. It clearly outlines the objective, input data format, and requirements, ensuring that the model understands the domain context. The prompt also includes a concrete example to set the expected output format.

The inputs to the LLM include technical indicators, sentiment scores, and basic stock features. The outputs are new alpha formulas, which are then computed into numerical values and used as additional features.

All LLM-generated formulaic alphas are normalized using Min–Max scaling before being fed into the prediction models. The scaling parameters are estimated from the training set and then applied to the testing set to avoid information leakage.

\subsection{Stock Prediction Using Transformer}
\begin{figure*}[!htbp]
    \centering
    \includegraphics[width=1.1\textwidth]{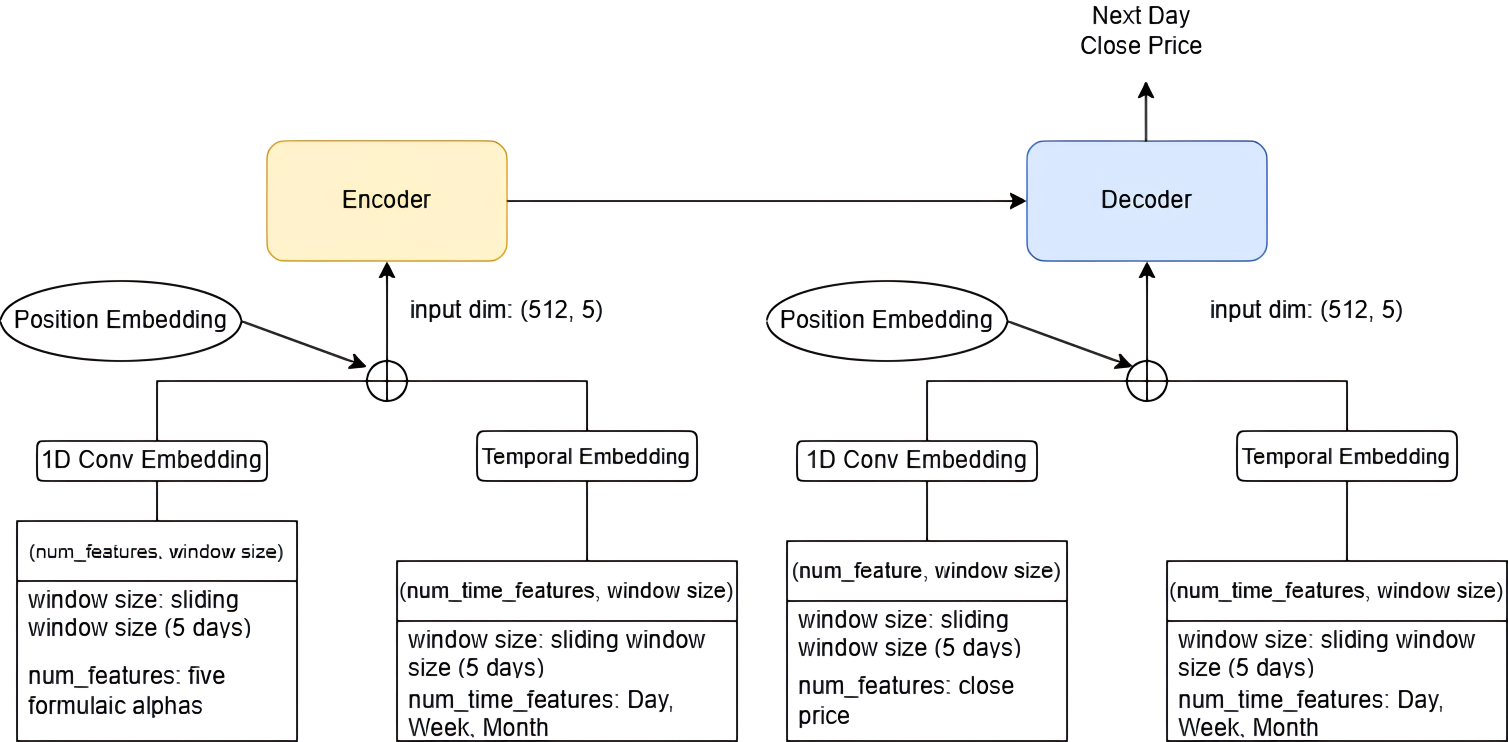}  
    \caption{Transformer Structure~\cite{chen2025sentiment}}
    \label{fig: transformer structure}
\end{figure*}

The features, the LLM-generated alphas, are fed into a vanilla Transformer model (Figure~\ref{fig: transformer structure}). The Transformer is designed to predict the closing price the next day for each stock.

Specifically, on the encoder side, the input includes five formulaic alphas with a sliding window of five days. These are passed through a 1D Convolutional Embedding layer to find useful patterns over time. At the same time, time-related features such as day, week, and month are handled by a Temporal Embedding layer to help the model understand time cycles. The outputs from these two layers are combined with Position Embedding, which helps the model remember the order of the data, and then sent into the Transformer Encoder. 

On the decoder side, the input is close prices with a sliding window of five days. These go through the same process: first a 1D Conv Embedding to learn patterns, then a Temporal Embedding for time features, and finally Position Embedding is added. The decoder uses both its own input and the encoder's output to make a prediction about the next day's closing price. This setup helps the model learn both price movements and time patterns to make better predictions.

The model is trained using the Mean Squared Error (MSE) loss function between the predicted and true closing prices. The model is optimized using the Adam optimizer with a learning rate of $5 \times 10^{-5}$ over 50 epochs. The batch size varies by stock, ranging from 3 to 64. Key hyperparameters, such as the learning rate and batch size, are selected through empirical tuning based on the training performance. All hyperparameter tuning was conducted exclusively on the training set. The test set was held out and used only for final evaluation. The adoption of a small batch size and a low learning rate serves to reduce the risk of overfitting. As demonstrated in prior studies~\cite{keskar2017largebatchtrainingdeeplearning,masters2018revisitingsmallbatchtraining, Peng2024CIFAR10}, these training constraints help the model capture generalizable patterns rather than idiosyncratic noise in the training data. A grid of commonly used values is explored, and the final configuration is chosen to balance predictive accuracy and training stability. Although the Transformer model was tuned using a simpler method, it still achieved competitive performance, indicating robustness against overfitting. Table~\ref{tab:hyperparameters} shows the final hyperparameters settings of the Transformer model.

\begin{table*}[htbp]
\centering
\caption{Final hyperparameter settings of the Transformer model}
\label{tab:hyperparameters}
\begin{tabular}{lc}
\hline
\textbf{Hyperparameter} & \textbf{Value} \\
\hline
Encoder layers  & 3 \\
Decoder layers   & 2 \\
Hidden size  & 512 \\
Feed-forward size  & 512 \\
Embedding size & 512 \\
Dropout rate & 0.0 \\
Activation function & GELU \\
Window length & 5 \\
\hline
\end{tabular}
\end{table*}

\subsection{Models Used for Performance Comparison}

In this study, some commonly used models in stock forecasting are added for comparison. The following is a brief description of each model used in this study. We employ Optuna~\cite{optuna_2019} to automatically search for suitable hyperparameter configurations for deep learning models and GridSearchCV for machine learning models such as XGBoost.

\begin{itemize}
    \item Long Short-Term Memory (LSTM) is a type of recurrent neural network that is good at learning patterns in time series data. It uses memory cells and gates to keep important information and forget less useful signals over time. In this study, the LSTM model employs three layers, each with a hidden dimension of 64.
    \item Temporal Convolutional Network (TCN) is a deep learning model that handles sequence data using 1D convolution layers. Unlike RNNs, it processes sequences in parallel and captures long-range dependencies using dilated convolutions. In this study, the TCN model consists of a single temporal block with 128 channels. Causal convolutions with a kernel size of 2 are used, and a dropout rate of 0.1 is applied.
    \item Support Vector Regression (SVR) is a machine learning method that finds a line or curve to fit the data while keeping errors within a set margin. It works well on small to medium-sized datasets and handles both linear and non-linear trends. In this study, the SVR uses a linear kernel with the regularization parameter C set to 10.
    \item Random Forest is an ensemble model that builds many decision trees and combines their results. It reduces overfitting and improves prediction accuracy by averaging the outcomes of individual trees. In this study, the Random Forest consists of 300 trees with a maximum tree depth of 30. The minimum number of samples required for node splitting and leaf nodes is set to 10 and 2, respectively.
    \item XGBoost (Extreme Gradient Boosting) is a tree-based machine learning method built on the idea of gradient boosting. It constructs an ensemble of decision trees in a sequential way, where each new tree focuses on correcting the errors made by previous trees. The model uses gradient information from the loss function to guide the learning process, which helps improve prediction accuracy. In this study, the XGBoost model is configured with a squared error objective, 100 boosting trees, a learning rate of 0.1, and a maximum tree depth of 3.
    \item Informer~\cite{zhou2021informerefficienttransformerlong} is a Transformer-based model designed for long time-series forecasting. It introduces a probabilistic sparse self-attention mechanism that selects the most relevant queries, which reduces computational cost while keeping important temporal information. By combining sparse attention with an efficient encoder–decoder structure, Informer can handle long input sequences and is suitable for large-scale forecasting tasks. In this study, the Informer model adopts a hidden dimension of 512 with eight attention heads and consists of three encoder layers and two decoder layers. ProbSparse attention is used to improve computational efficiency, along with fixed time embeddings. The remaining settings follow standard configurations, including GELU activation and zero dropout.
\end{itemize}

\subsection{Evaluation}

MSE is used to evaluate the accuracy of the stock price predictions generated by the models. It measures the average squared difference between the predicted stock prices and the actual prices. A lower MSE indicates that the predicted values are closer to the true values, which means better prediction accuracy. The formula for MSE is the following.
    
\[
\text{MSE} = \frac{1}{n} \sum_{i=1}^{n} (y_i - \hat{y}_i)^2, 
\]

\noindent where \(y_i\) is the actual stock price, \(\hat{y}_i\) is the predicted stock price, and \(n\) is the total number of predictions.

Furthermore, to highlight the novelty of incorporating sentiments from both target companies and their related firms, we evaluate the impact of integrating sentiment information into a Transformer model for stock price forecasting, which will be illustrated in Section~\ref{Sec 4.3}. For both the baseline scenario without sentiment and the experimental scenario with sentiment, we conduct ten independent runs per stock. This number is chosen as a common practice in empirical machine learning research to balance statistical robustness with computational efficiency. Furthermore, to examine the feature randomness and training variance, two experimental setups are considered: (i) fixing a single alpha set and repeating the experiment of all models for 10 runs to account for training variance; and (ii) fixing the training seed of the Transformer model and generating ten different alpha sets to account for feature randomness. The experimental setup (ii)  includes the comparison between Transformer trained with and without sentiment information. 

We report the average MSE and the standard deviation in each case. This ensures a fair and statistically sound comparison, allowing us to assess not only the average performance improvement of sentiment features, but also the consistency of the model predictions across multiple runs.

\subsubsection{Diebold--Mariano Test}
In addition to error-based metrics, we employ the Diebold--Mariano (DM) test~\cite{4078c4a6-012f-36a3-8447-1b02ae8abba2} to assess whether the forecasting performance differences between models are statistically significant. Consider two competing forecasting models, denoted as model $A$ and model $B$. Let $y_t$ be the observed stock price at time $t$, and $\hat{y}_{t}^{A}$ and $\hat{y}_{t}^{B}$ be the corresponding out-of-sample forecasts. For a given loss function $L(\cdot)$, the loss differential is defined as
\begin{equation}
d_t = L(y_t, \hat{y}_{t}^{A}) - L(y_t, \hat{y}_{t}^{B}).
\end{equation}
In this study, we adopt the mean squared error as the loss function, i.e., $L(y_t, \hat{y}_t) = (y_t - \hat{y}_t)^2$.

The DM test examines the null hypothesis that the two models have equal predictive accuracy, $H_0: \mathbb{E}[d_t] = 0$. The test statistic is calculated as $S = \bar{d} / \sqrt{\hat{V}(\bar{d})}$, where $\bar{d}$ is the sample mean of the loss differential and $\hat{V}(\bar{d})$ is an estimate of its variance accounting for serial correlation.

To account for potential bias in small sample sizes and for the prediction horizon $h$, we apply the Harvey-Leybourne-Newbold (HLN) correction~\cite{HARVEY1997281}. The corrected DM statistic, $S^*$, is computed as follows:
\begin{equation}
S^* = S \cdot \sqrt{\frac{T + 1 - 2h + h(h - 1)/T}{T}},
\end{equation}
where $T$ represents the number of observations in the out-of-sample period and $h$ is the forecast horizon. In our empirical evaluation, we set $h=1$ for one-step-ahead predictions. The corrected statistic $S^*$ is then compared against the critical values of a Student's $t$-distribution with $(T-1)$ degrees of freedom. A statistically significant $S^*$ indicates that the difference in forecasting accuracy is unlikely to be driven by random variation.

Finally, since we perform multiple pairwise comparisons between different models, there is a higher risk of finding significant results by pure chance. To fix this, we apply the Holm-Bonferroni adjustment~\cite{29def780-e117-38f0-8afb-edf384af3fad} to our $p$-values. This method works by ranking the $p$-values from smallest to largest and adjusting each one with a different weight. This ensures that our statistical conclusions remain strict and reliable even when many tests are conducted at the same time.

\section{Results}\label{result}

Table~\ref{tab: alphas of five companies} shows the LLM-generated formulaic alphas for all the five companies. Table~\ref{tab: sentiment_comparison} demonstrates the Transformer performance with and without sentiment scores as input. Figure~\ref{fig: llm_reasoning_process} shows the reasoning process of the LLM to generate the formulaic alphas for Toyota. Table~\ref{tab:mse_comparison} shows the prediction performance in terms of MSE for all companies using the formulaic alphas in Table~\ref{tab: alphas of five companies}. Figure~\ref{fig:model_comparison} visualizes the prediction performances of some models for HSBC stock prediction. Based on the figures, the models are able to capture stock price movements when LLM-generated features are used as inputs. More details of the results will be illustrated in the subsequent subsections.

\subsection{Alpha Generation}

\begin{sidewaysfigure}
    \centering
    \includegraphics[width=\textheight]{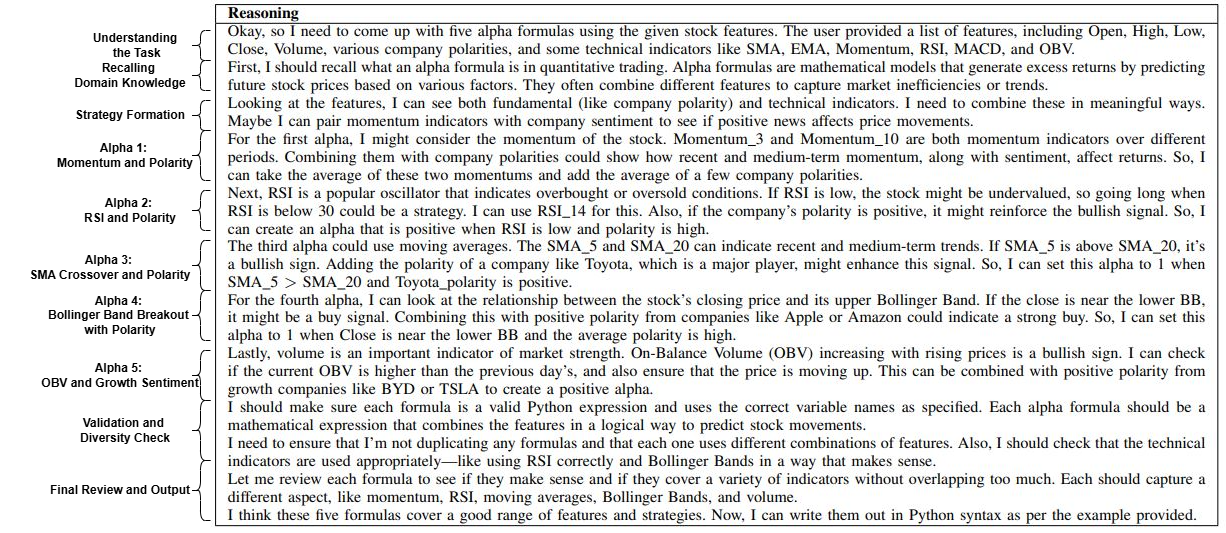}
    \caption{Reasoning Process for Generating Toyota's Alpha}
    \label{fig: llm_reasoning_process}
\end{sidewaysfigure}

According to Figure~\ref{fig: llm_reasoning_process}, the DeepSeek model generates formulaic alphas by emulating a human-like reasoning process that systematically integrates both technical indicators and sentiment-based features. The reasoning process is automatically generated by the LLM. Specifically, the model begins by recalling the definition and purpose of alpha formulas in quantitative trading. It then strategically combines diverse inputs such as momentum indicators (e.g., Momentum\_3, Momentum\_10), oscillators (e.g., RSI\_14), trend-following metrics (e.g., SMA\_5, SMA\_20), and volume-based measures (e.g., OBV), alongside company-specific polarity scores derived from sentiment analysis. Each formula is crafted to capture a distinct market dynamic, such as momentum-sentiment alignment, oversold conditions reinforced by positive sentiment, or price movements relative to Bollinger Bands, to ensure variety and avoid redundancy. 

Through this reasoning process, the DeepSeek model enables analysts and traders to understand the rationale behind each predictive signal rather than relying on black-box outputs. This transparency supports better risk assessment, as users can evaluate the logic and market conditions each alpha captures. Moreover, by systematically combining diverse indicators, the process encourages the discovery of non-obvious relationships and helps build more robust, diversified trading strategies. For example, for Pepsi (in Table~\ref{tab: alphas of five companies}), one non-obvious relationship captured by the LLM appears in $\alpha_1$: 

\[
\alpha_{1,t} = \text{Momentum}_{3,t} + \text{Amazon\_polarity}_t.
\]

Although Amazon and Pepsi belong to different sectors, e-commerce and consumer beverages, respectively, the model identifies a subtle link. The inclusion of \texttt{Amazon\_polarity\_t} suggests that positive sentiment toward Amazon may reflect broader consumer optimism or spending strength, which, in turn, can have downstream effects on consumer goods companies such as Pepsi. Such a relationship is not immediately intuitive, which reveals the ability of LLM to infer macroeconomic or behavioral connections across sectors based on sentiment signals.

Although such an alpha is not guaranteed to contain strong predictive power on its own, it serves an important role in expanding the space of candidate alphas. 


\subsection{Stock Prediction Performance}

Table~\ref{tab:mse_comparison} reports the stock price prediction performance of different models measured by MSE. The comparison is conducted before inverse transformation, as our primary interest lies in the relative performance across models rather than the absolute price scale.

\begin{figure*}[!htbp]
  \centering

  \begin{subfigure}{0.3\textwidth}
    \centering
    \includegraphics[width=\linewidth]{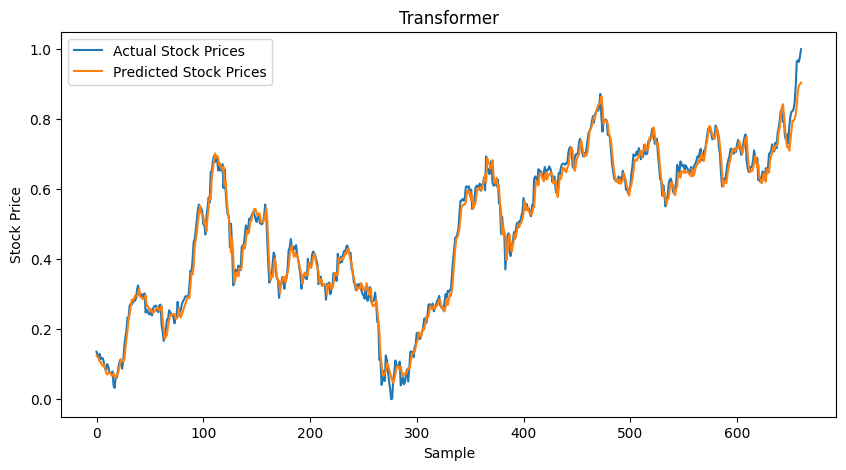}
    \caption{Transformer}
  \end{subfigure}
  \hfill
  \begin{subfigure}{0.3\textwidth}
    \centering
    \includegraphics[width=\linewidth]{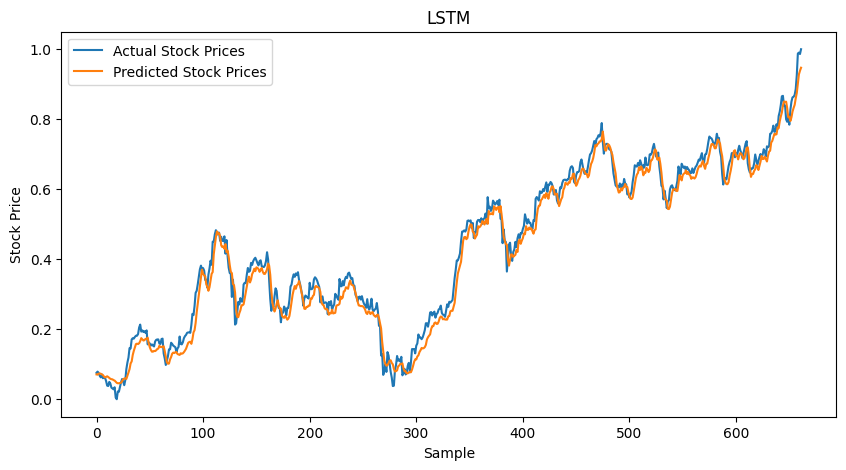}
    \caption{LSTM}
  \end{subfigure}
  \hfill
  \begin{subfigure}{0.3\textwidth}
    \centering
    \includegraphics[width=\linewidth]{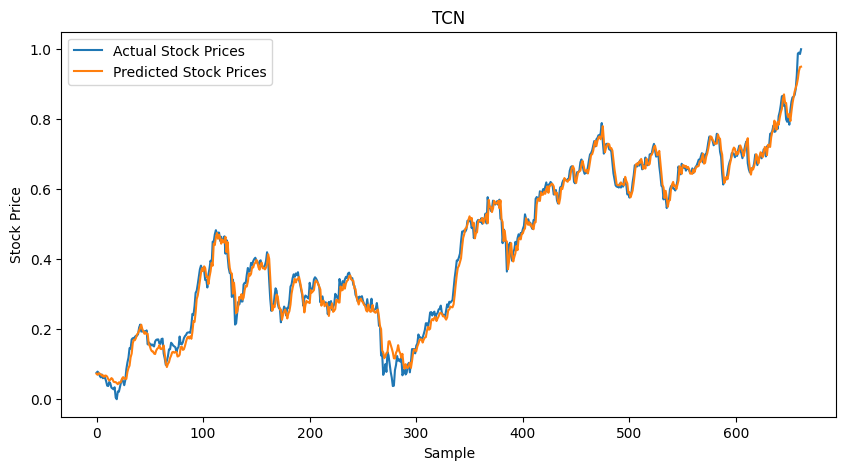}
    \caption{TCN}
  \end{subfigure}

  \vspace{0.5cm}

  \begin{subfigure}{0.3\textwidth}
    \centering
    \includegraphics[width=\linewidth]{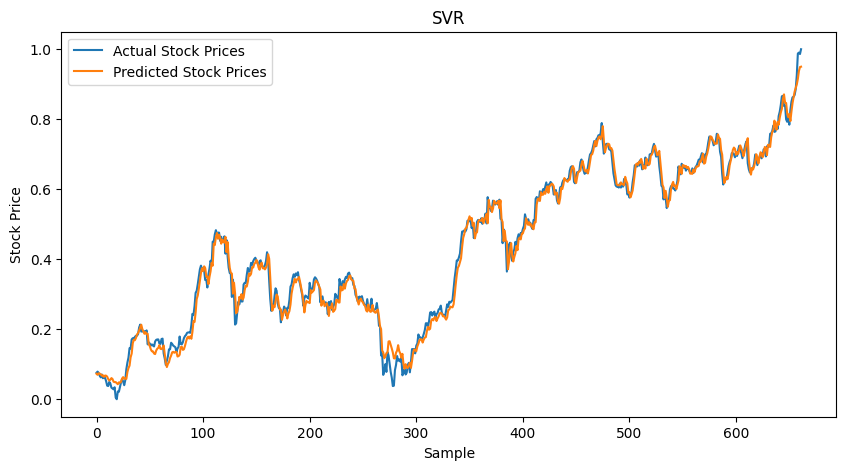}
    \caption{SVR}
  \end{subfigure}
  \hspace{0.05\textwidth} 
  \begin{subfigure}{0.3\textwidth}
    \centering
    \includegraphics[width=\linewidth]{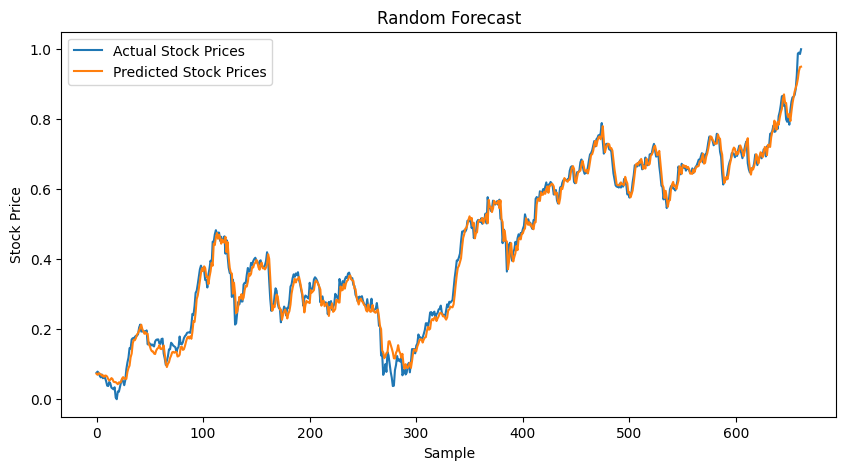}
    \caption{Random Forest}
  \end{subfigure}

  \caption{Comparison of prediction results from different models}
  \label{fig:model_comparison}
\end{figure*}

According to Table~\ref{tab:mse_comparison}, for most of the time, Transformer has the best performance in terms of MSE, followed by Informer. The plotted graph for the Transformer (Figure~\ref{fig:model_comparison}) shows that the predicted stock prices closely follow the actual stock price trend, indicating its strong ability to capture underlying patterns in the data. In addition, the strong performance of the Transformer and Informer models highlights the advantages of attention mechanisms in sequential modeling.

LSTM also performs well, though not as accurately as the Transformer. Its predictions generally follow the actual price movements, but with slightly more noticeable deviations. Similarly, the TCN model manages to capture the general shape of the stock price trends. For some time, it can outperform LSTM.

For traditional machine learning models, the SVR has the worst performance. The ensemble methods, random forest and XGBoost, can sometimes generate competitive performance compared to the Transformer (e.g., HSBC and Toyota).

Table~\ref{tab:dm_summary_all} reports the overall Diebold--Mariano (DM) test results comparing the Transformer with the baseline models across all stocks. The statistical outcomes are consistent with the MSE rankings. In particular, the Transformer shows significantly better forecasting accuracy than LSTM, TCN, SVR, and the ensemble models, with all differences being statistically significant ($p < 0.01$). In contrast, the difference between the Transformer and Informer is not statistically significant ($p > 0.05$), suggesting that their similar attention-based structures lead to comparable predictive performance.

To provide a clearer view of the pairwise model comparisons, Table~\ref{tab:dm_test} reports the Diebold--Mariano test results for Pepsi stock. This stock is used as a representative example to illustrate the relative performance differences between models. The results show that the Transformer achieves statistically better forecasting accuracy than most baseline models. Similar patterns are observed for the other stocks in our experiments.

Overall, the analysis highlights the advantage of Transformer-based architectures in modeling financial time series, especially when supported by informative signals generated from large language models. Furthermore, the LLM-generated formulaic features have a strong predictive ability. We further compare the results with other alphas (featuretools-generated alphas and human-defined alphas) in Section~\ref{Sec5}. 

In addition, the reasoning process of LLMs in generating formulaic alphas can serve as a valuable reference to improve financial decision making. Furthermore, the novel insights provided by LLMs can inspire human investors to develop new features that improve the prediction of stock prices.

\begin{table*}[htbp]
\centering
\caption{Alpha Formulas for Five Companies}
\label{tab: alphas of five companies}
\begin{tabular}{@{}p{0.15\linewidth}p{0.92\linewidth}@{}}
\toprule
\textbf{Company} & \textbf{Alpha Formulas} \\
\midrule

\textbf{Apple} &

\begin{minipage}[t]{\linewidth}
\scriptsize
\begin{lstlisting}[language=Python, basicstyle=\footnotesize\ttfamily]
alpha1_t = (C_t - O_t) / O_t + 0.5 * (Apple_polarity_t + 
           Google_polarity_t)
alpha2_t = (High_t - Low_t) / Low_t + 0.3 * (Amazon_polarity_t                
           - Microsoft_polarity_t)
alpha3_t = (C_t - SMA_5_t) / SMA_5_t + 0.2 * (TSLA_polarity_t 
            + Nvidia_polarity_t)
alpha4_t = (Momentum_3_t + Momentum_10_t) / 2 + 0.4 * (
            Meta_polarity_t - Intel_polarity_t)
alpha5_t = (RSI_14_t - 50) / 50 + 0.6 * (AMD_polarity_t + 
            IBM_polarity_t)
\end{lstlisting} 
\end{minipage}
\\

\textbf{HSBC} &
\begin{minipage}[t]{\linewidth}
\scriptsize
\begin{lstlisting}[language=Python, basicstyle=\footnotesize\ttfamily]
alpha1_t = (C_t - O_t) / O_t + 0.3 * Apple_polarity_t + 0.2 * (SMA_5_t - 
           SMA_20_t)
alpha2_t = (RSI_14_t / 14) * Momentum_10_t + 0.4 * UBS_polarity_t - 
            0.2 * BB_Lower_t
alpha3_t = (MACD_t - MACD_Signal_t) + 0.5 * Microsoft_polarity_t + 0.2 * 
           (V_t / V_{t-1})
alpha4_t = (C_t / SMA_20_t) - 1 + 0.3 * Morgan_Stanley_polarity_t - 0.1 * 
           (BB_Upper_t - BB_Lower_t)
alpha5_t = (Momentum_3_t + Momentum_10_t) / 2 + 0.4 * Goldman_Sachs_polarity_t 
           - 0.2 * OBV_t
\end{lstlisting} 
\end{minipage}
\\

\textbf{Pepsi} &
\begin{minipage}[t]{\linewidth}
\scriptsize
\begin{lstlisting}[language=Python, basicstyle=\footnotesize\ttfamily]
alpha1_t = Momentum_3_t + Amazon_polarity_t
alpha2_t = (SMA_5_t - SMA_20_t) + Microsoft_polarity_t
alpha3_t = (V_t / V_{t-1}) * Walmart_polarity_t
alpha4_t = (MACD_t - MACD_Signal_t) + IBM_polarity_t
alpha5_t = (50 - RSI_14_t) + TSLA_polarity_t
\end{lstlisting} 
\end{minipage}
\\

\textbf{Tencent} &
\begin{minipage}[t]{\linewidth}
\scriptsize
\begin{lstlisting}[language=Python, basicstyle=\footnotesize\ttfamily]
alpha1_t = (Momentum_3_t + Momentum_10_t) / 2 + 0.3 * (V_t / V_{t-1}) + 
            0.2 * polarity_avg
alpha2_t = (SMA_5_t - SMA_20_t) + 0.5 * (C_t - O_t) / O_t + 0.4 * polarity_avg
alpha3_t = (RSI_14_t - 50) / 50 + 0.6 * (OBV_t - OBV_{t-20}) / OBV_{t-20} 
           + 0.3 * polarity_avg
alpha4_t = (Momentum_3_t + Momentum_10_t) / 2 + 0.4 * (EMA_10_t - EMA_10_{t-1}) 
           + 0.2 * polarity_avg
alpha5_t = (MACD_t - MACD_Signal_t) + 0.5 * (BB_Upper_t - BB_Lower_t) 
           / BB_Lower_t + 0.3 * polarity_avg
# polarity_avg = mean of 12 tech company polarities
\end{lstlisting} 
\end{minipage}
\\

\textbf{Toyota} &
\begin{minipage}[t]{\linewidth}
\scriptsize
\begin{lstlisting}[language=Python, basicstyle=\footnotesize\ttfamily]
alpha1_t = (Momentum_3_t + Momentum_10_t) / 2 + (Toyota_polarity_t + 
           Nissan_polarity_t + Honda_polarity_t) / 3
alpha2_t = RSI_14_t * (1 if RSI_14_t < 30 else 0) * 
           (1 if Toyota_polarity_t > 0 else 0)
alpha3_t = 1 if (SMA_5_t > SMA_20_t) and (Toyota_polarity_t > 0) else 0
alpha4_t = (C_t - BB_Lower_t) / (BB_Upper_t - BB_Lower_t) * 
           (1 if Apple_polarity_t > 0 and Amazon_polarity_t > 0 else 0)
alpha5_t = (OBV_t > OBV_{t-1}) * (C_t > C_{t-1}) * (1 if BYD_polarity_t > 0 
           and TSLA_polarity_t > 0 else 0)
\end{lstlisting} 
\end{minipage}
\\

\bottomrule
\end{tabular}
\end{table*}

\begin{table*}[htbp]
\centering
\caption{MSE Comparison of Models for Stock Price Prediction with LLM-Generated Alphas, with lowest MSE in each column in bold and second-lowest underlined.}
\begin{tabular}{@{}lccccc@{}}
\toprule
\textbf{Model} & \textbf{Apple} & \textbf{HSBC} & \textbf{Pepsi} & \textbf{Tencent} & \textbf{Toyota} \\
\midrule
Transformer    
& \textbf{0.0003} 
& \textbf{0.0004} 
& \textbf{0.0004} 
& \textbf{0.0002} 
& \textbf{0.0001} \\

Informer        
& \textbf{0.0003} 
& 0.0007 
& \underline{0.0006} 
& \underline{0.0004} 
& \underline{0.0002} \\

LSTM           
& 0.0030 
& 0.0009 
& 0.0018 
& 0.0022 
& 0.0015 \\

TCN            
& \underline{0.0018} 
& 0.0006 
& 0.0028 
& 0.0077 
& 0.0046 \\

SVR            
& 0.0048 
& 0.0024 
& 0.0046 
& 0.0052 
& 0.0048 \\

Random Forest  
& 0.0019 
& \textbf{0.0004} 
& 0.0014 
& 0.0011 
& 0.0005 \\

XGBoost        
& 0.0019 
& \underline{0.0005}
& 0.0014 
& 0.0011 
& 0.0005 \\
\bottomrule
\end{tabular}
\label{tab:mse_comparison}
\end{table*}

\begin{table*}[htbp]
\centering
\caption{Diebold--Mariano Test Summary: Transformer vs. Baseline Models Across All Stocks. Each cell reports the DM statistic with $p$-value in parentheses. A negative statistic indicates the Transformer's superior predictive accuracy.}
\label{tab:dm_summary_all}
\begin{adjustbox}{max width=\textwidth}
\begin{tabular}{lccccc}
\toprule
\textbf{Baseline Model} & \textbf{Apple} & \textbf{HSBC} & \textbf{Pepsi} & \textbf{Tencent} & \textbf{Toyota} \\
\midrule
Informer        & $-0.27$ (0.787) & $-0.48$ (0.631) & $-0.64$ (0.524) & $-0.39$ (0.697) & $-0.71$ (0.478) \\
LSTM            & $-20.08$***   & $-16.75$***   & $-7.97$***   & $-12.31$***   & $-8.05$***   \\
TCN             & $-29.63$***   & $-12.56$***   & $-10.62$***  & $-16.26$***  & $-9.88$***   \\
SVR             & $-16.56$***  & $-19.63$***  & $-16.02$***  & $-25.84$***  & $-14.76$***  \\
Random Forest   & $-8.44$***   & $-13.27$***   & $-5.34$***   & $-10.53$***   & $-5.12$***   \\
XGBoost         & $-8.81$***   & $-13.44$***   & $-5.76$***   & $-10.62$***   & $-5.44$***   \\
\bottomrule
\end{tabular}
\end{adjustbox}
\begin{center}
\footnotesize{Note: (***) denotes $p < 0.01$. Negative DM values indicate lower loss for the Transformer model.}
\end{center}
\end{table*}

\begin{sidewaystable*}[htbp]
\centering
\renewcommand{\arraystretch}{1.25}
\caption{Diebold--Mariano test results for Pepsi stock.
Each cell compares the \emph{row model} against the \emph{column model}.
The test statistic is reported with the corresponding Holm–Bonferroni adjusted $p$-value shown in parentheses.
The loss differential is defined as $d_t = L_{\text{row}} - L_{\text{column}}$,
such that a negative and statistically significant DM statistic indicates superior predictive accuracy of the row model.
Statistical significance is denoted by *** ($p<0.01$), ** ($p<0.05$), and * ($p<0.10$).}
\label{tab:dm_test}

\begin{adjustbox}{max width=\textheight}
\begin{tabular}{lccccccc}
\toprule
\textbf{Row model $\downarrow$ \textbackslash\ Column model $\rightarrow$}
& \textbf{Transformer}
& \textbf{Informer}
& \textbf{LSTM}
& \textbf{TCN}
& \textbf{SVR}
& \textbf{Random Forest}
& \textbf{XGBoost} \\
\midrule
Transformer
& --
& \makecell{$-0.64$ \\ (0.5237)}
& \makecell{$-7.97$*** \\ (6.90$\times 10^{-14}$)}
& \makecell{$-10.62$*** \\ (2.45$\times 10^{-23}$)}
& \makecell{$-16.02$*** \\ (6.94$\times 10^{-48}$)}
& \makecell{$-5.34$*** \\ (3.84$\times 10^{-7}$)}
& \makecell{$-5.76$*** \\ (6.20$\times 10^{-8}$)} \\
Informer
&
& --
& \makecell{$-8.17$*** \\ (1.74$\times 10^{-14}$)}
& \makecell{$-10.62$*** \\ (2.45$\times 10^{-23}$)}
& \makecell{$-16.06$*** \\ (4.64$\times 10^{-48}$)}
& \makecell{$-5.39$*** \\ (3.84$\times 10^{-7}$)}
& \makecell{$-5.83$*** \\ (6.06$\times 10^{-8}$)} \\
LSTM
&
&
& --
& \makecell{$-7.78$*** \\ (2.45$\times 10^{-13}$)}
& \makecell{$-13.72$*** \\ (9.83$\times 10^{-37}$)}
&  \makecell{$-6.54$*** \\ (1.00$\times 10^{-9}$)}
& \makecell{$-5.81$*** \\ (6.05$\times 10^{-8}$)} \\
TCN
&
&
&
& --
& \makecell{$-9.83$*** \\ (2.45$\times 10^{-20}$)}
& \makecell{11.19*** \\ (1.27$\times 10^{-25}$)}
& \makecell{11.60*** \\ (2.85$\times 10^{-27}$)}\\
SVR
&
&
&
&
& --
& \makecell{15.39*** \\ (9.48$\times 10^{-45}$)}
& \makecell{15.37*** \\ (1.02$\times 10^{-44}$)}\\
Random Forest
&
&
&
&
&
& --
& \makecell{$-1.18$ \\ (0.4734)}\\
XGBoost
&
&
&
&
&
&
& -- \\
\bottomrule
\end{tabular}
\end{adjustbox}
\end{sidewaystable*}

\subsection{Effect of Sentiment Integration on Prediction Accuracy} \label{Sec 4.3}

Table~\ref{tab: sentiment_comparison} compares the performance of the Transformer model with and without sentiment information. The experiments are run ten times for each stock to consider the feature randomness because each run generates a new alpha set. Across the five companies, the incorporation of sentiment features leads to noticeable improvements in prediction accuracy, as reflected in consistently lower MSE. Additionally, standard deviations are generally reduced in the sentiment-enhanced setting, suggesting more stable and robust model behavior. This indicates that sentiment information not only contributes to better predictive performance, but also helps mitigate variability across different runs. Although the magnitude of improvement varies by company, the consistent trend reinforces the value of leveraging sentiment from both target firms and their related entities in stock price forecasting.

Beyond average performance, standard deviation results provide insight into the stability of LLM-generated formulaic alphas. Despite the fact that each training run involves a newly generated set of alpha signals from the LLM, the relatively low standard deviations across all stocks demonstrate that the model consistently learns meaningful patterns. This suggests that the LLM-generated alphas possess strong generalization properties, allowing the Transformer to maintain stable performance even under varying input conditions. Such robustness is particularly valuable in financial applications, where consistency between repeated trials is essential for real-world deployment.

\begin{table}[htbp]
    \centering
    \caption{Transformer performance with and without sentiment information across 10 different alpha sets, reported as mean and standard deviation of MSE}
    \label{tab: sentiment_comparison}
    \begin{tabular}{lcc}
        \toprule
        Company & With Sentiment & Without Sentiment \\
        \midrule
        Apple   & 0.00029 ($1.38 \times 10^{-4}$) & 0.00034 ($1.56 \times 10^{-4}$) \\
        HSBC    & 0.00040 ($2.35 \times 10^{-4}$) & 0.00044 ($2.43 \times 10^{-4}$) \\
        Pepsi   & 0.00038 ($0.87 \times 10^{-4}$) & 0.00041 ($0.94 \times 10^{-4}$) \\
        Tencent & 0.00050 ($2.88 \times 10^{-4}$) & 0.00052 ($3.12 \times 10^{-4}$) \\
        Toyota  & 0.00014 ($0.91 \times 10^{-4}$) & 0.00019 ($0.96 \times 10^{-4}$) \\
        \bottomrule
    \end{tabular}
\end{table}

\begin{table*}[htbp]
\centering
\caption{MSE comparison of stock price prediction models using a fixed alpha set over 10 runs (mean and standard deviation), with the best (lowest MSE) in each column in bold and the second best underlined.}
\begin{tabular}{@{}lccccc@{}}
\toprule
\textbf{Model} & \textbf{Apple} & \textbf{HSBC} & \textbf{Pepsi} & \textbf{Tencent} & \textbf{Toyota} \\
\midrule
Transformer &
\makecell{\underline{0.00056} \\ {\footnotesize($3.04 \times 10^{-4}$)}} &
\makecell{\textbf{0.00040} \\ {\footnotesize($1.72 \times 10^{-4}$)}} &
\makecell{\textbf{0.00020} \\ {\footnotesize($1.32 \times 10^{-4}$)}} &
\makecell{\textbf{0.00047} \\ {\footnotesize($4.06 \times 10^{-4}$)}} &
\makecell{\textbf{0.00049} \\ {\footnotesize($1.84 \times 10^{-4}$)}} \\

Informer &
\makecell{\textbf{0.00054} \\ {\footnotesize($2.82 \times 10^{-4}$)}} &
\makecell{0.00074 \\ {\footnotesize($2.54 \times 10^{-4}$)}} &
\makecell{\underline{0.00046} \\ {\footnotesize($2.37 \times 10^{-4}$)}} &
\makecell{\underline{0.00094} \\ {\footnotesize($3.35 \times 10^{-4}$)}} &
\makecell{0.00052 \\ {\footnotesize($2.24 \times 10^{-4}$)}} \\

LSTM &
\makecell{0.00298 \\ {\footnotesize($1.06 \times 10^{-4}$)}} &
\makecell{0.00071 \\ {\footnotesize($1.48 \times 10^{-4}$)}} &
\makecell{0.00170 \\ {\footnotesize($2.47 \times 10^{-4}$)}} &
\makecell{0.00172 \\ {\footnotesize($1.22 \times 10^{-4}$)}} &
\makecell{0.00128 \\ {\footnotesize($3.98 \times 10^{-4}$)}} \\

TCN &
\makecell{0.00178 \\ {\footnotesize($1.03 \times 10^{-4}$)}} &
\makecell{0.00054 \\ {\footnotesize($0.35 \times 10^{-4}$)}} &
\makecell{0.00266 \\ {\footnotesize($5.39 \times 10^{-4}$)}} &
\makecell{0.00186 \\ {\footnotesize($3.99 \times 10^{-4}$)}} &
\makecell{0.00302 \\ {\footnotesize($6.74 \times 10^{-4}$)}} \\

SVR &
0.00479 &
0.00157 &
0.00436 &
0.00251 &
0.00481 \\

Random Forest &
\makecell{0.00186 \\ {\footnotesize($0.09 \times 10^{-4}$)}} &
\makecell{\underline{0.00044} \\ {\footnotesize($0.02 \times 10^{-4}$)}} &
\makecell{0.00140 \\ {\footnotesize($0.15 \times 10^{-4}$)}} &
\makecell{0.00103 \\ {\footnotesize($0.05 \times 10^{-4}$)}} &
\makecell{\underline{0.00050} \\ {\footnotesize($0.05 \times 10^{-4}$)}} \\

XGBoost &
\makecell{0.00189 \\ {\footnotesize($0.72 \times 10^{-4}$)}} &
\makecell{0.00049 \\ {\footnotesize($0.21 \times 10^{-4}$)}} &
\makecell{0.00151 \\ {\footnotesize($0.49 \times 10^{-4}$)}} &
\makecell{0.00116 \\ {\footnotesize($0.29 \times 10^{-4}$)}} &
\makecell{0.00051 \\ {\footnotesize($0.15 \times 10^{-4}$)}} \\
\bottomrule
\end{tabular}
\label{tab:training_variance}
\end{table*}

\begin{table*}[htbp]
\centering
\caption{Variance decomposition of MSE across repeated experiments based on results of Table~\ref{tab:training_variance}.
The total variance is decomposed into between-model variance and within-model variance based on 10 independent runs for each stock.
The percentage columns indicate the relative contribution of each component to the total variance.}
\label{tab:variance_decomposition}
\begin{adjustbox}{max width=\textwidth}
\begin{tabular}{lccccc}
\toprule
\textbf{Stock} 
& \textbf{Between-Model Variance} 
& \textbf{Within-Model Variance} 
& \textbf{Total Variance} 
& \textbf{Between (\%)} 
& \textbf{Within (\%)} \\
\midrule
Apple   & $2.17 \times 10^{-6}$ & $3.32 \times 10^{-8}$ & $2.20 \times 10^{-6}$ & 98.49 & 1.51 \\
HSBC    & $1.64 \times 10^{-7}$ & $1.96 \times 10^{-8}$ & $1.84 \times 10^{-7}$ & 89.34 & 10.66 \\
Pepsi   & $1.98 \times 10^{-6}$ & $7.13 \times 10^{-8}$ & $2.05 \times 10^{-6}$ & 96.53 & 3.47 \\
Tencent & $4.69 \times 10^{-7}$ & $7.53 \times 10^{-8}$ & $5.45 \times 10^{-7}$ & 86.17 & 13.83 \\
Toyota  & $2.87 \times 10^{-6}$ & $1.16 \times 10^{-7}$ & $2.99 \times 10^{-6}$ & 96.11 & 3.89 \\
\bottomrule
\end{tabular}
\end{adjustbox}
\end{table*}

\subsection{Training Stability of Models with a Fixed Alpha Set}

Table~\ref{tab:training_variance} illustrates the performance comparison of different models using a fixed alpha set over ten runs, reported as mean and standard deviation. In this study, SVR adopts a linear kernel setup, so the results of SVR are deterministic.

The results show clear differences across models when a fixed alpha set is used. Overall, attention-based models deliver more accurate predictions than the other approaches. The Transformer model performs consistently well across most stocks, suggesting that its attention structure is effective in capturing time dependencies under this setting. Informer also shows competitive performance, especially for companies such as Apple, but its results are relatively less stable across different stocks.

In contrast, recurrent and convolution-based models, such as LSTM and TCN, tend to produce higher prediction errors. This indicates that these models may have difficulty modeling long-range patterns between different alphas. Traditional machine learning methods show mixed behavior. Tree-based models such as Random Forest and XGBoost perform reasonably well for some stocks but lack consistency across all cases, while SVR generally yields weaker results.

These findings suggest that attention-based architectures are more suitable for stock price forecasting when the same alpha set is applied across models. Their ability to focus on relevant time steps likely helps improve both accuracy and stability, whereas other models appear more sensitive to feature limitations.

Besides, to further examine the robustness of the empirical results, we conduct a variance decomposition analysis based on repeated experiments.
Specifically, the total variance of the MSE is decomposed into two components: 
(i) between-model variance, which captures performance differences attributable to model choice, and 
(ii) within-model variance, which reflects variability arising from training randomness such as parameter initialization and stochastic optimization.
This analysis allows us to assess whether the observed performance differences are primarily driven by model selection or by random fluctuations across runs.

Table~\ref{tab:variance_decomposition} presents the variance decomposition results for all five stocks.
Across all cases, the between-model variance accounts for the dominant share of the total variance, ranging from 86.17\% to 98.49\%.
In contrast, the within-model variance remains relatively small, indicating that the forecasting performance of each model is stable across repeated runs.
These findings suggest that the differences observed in the MSE comparison are mainly driven by inherent model characteristics rather than training randomness.

\section{Further Discussion} \label{Sec5}

\subsection{Comparison of Different Feature Generation Approaches}

\subsubsection{\textbf{Featuretools-Generated Features}}

To highlight the advantages of using LLMs for feature generation, several methods are compared. One of them is Featuretools~\cite{kanter2015deep}, an automatic feature engineering tool. It creates new features by applying transformation and aggregation operations to the original data. For example, it can combine price and sentiment data using basic math operations such as addition, subtraction, or division. 

Table~\ref{tab:features generated by featuretools} shows the top five features per company generated by Featuretools based on Predictive Power Score (PPS)~\cite{florian_wetschoreck_2020_4091345}. PPS is a score between 0 and 1 that captures both linear and non-linear relationships. For each company, we calculate the PPS between all featuretool-generated alphas and the close price, and select the top five alphas with the highest scores. These selected alphas are then used as input features for the prediction models.

\begin{table*}[ht]
\centering
\caption{Top Five Features per Company Generated by Featuretools (Based on Predictive Power Score)}
\label{tab:features generated by featuretools}
\resizebox{\textwidth}{!}{ 
\begin{tabular}{lccccc}
\hline
\textbf{Feature} & \textbf{Apple} & \textbf{HSBC} & \textbf{Pepsi} & \textbf{Tencent} & \textbf{Toyota} \\
\hline
Feature 1 & Close - Google\_polarity & ABSOLUTE(Close) & PERCENTILE(Close) &  -(Close) & Close + Honda\_polarity \\
Feature 2 & Close - Intel\_polarity & -(Close) & Close + Walmart\_polarity & Apple\_polarity - Close & BMW\_polarity + Close \\
Feature 3 & Close - TSLA\_polarity & 1 / Close & 1 / Close & ABSOLUTE(Close) & BYD\_polarity + Close \\
Feature 4 & Close + Google\_polarity & PERCENTILE(Close) & Close - Colgate\_polarity & SQUARE\_ROOT(Close) & Close - Ford\_polarity \\
Feature 5 & Close + Samsung\_polarity & SQUARE\_ROOT(Close) & -(Close) & Close + Tencent\_polarity & Close - GM\_polarity \\
\hline
\end{tabular}%
}
\end{table*}

According to Table~\ref{tab:features generated by featuretools}, the drawbacks of this method are obvious. First of all, the generated features are mathematically valid, but they are hard to interpret. For example, it is hard to explain the meaning of "Close - TSLA\_polarity" in decision-making. In addition, some generated features are redundant and noisy. For example, "Close + Google\_polarity" and "Close - Google\_polarity" are highly correlated, which may confuse the model and reduce generalization. Furthermore, the generated features lack diversity, as most of the formulas follow a similar structure.

Table~\ref{tab:combined_mse_comparison} shows the comparison of the stock forecast performance of the Featuretools-generated alpha with the LLM-generated performance. In general, the Featuretools-generated alphas yield less accurate results compared to the LLM-generated ones.

\subsubsection{\textbf{Human-Defined Formulaic Alphas}}

Another method used for comparison is human-defined formulaic alphas. These features are taken from the paper \textit{"101 Formulaic Alphas"} by Zura Kakushadze~\cite{kakushadze2016101formulaicalphas}. This paper is well-known in quantitative finance and is widely cited in both academic research and industry practice. It provides a collection of 101 trading formulas that were developed using real trading experience and historical market behavior. The alphas were designed at a quantitative hedge fund and are based on price, volume, and other market variables. Each formula is manually crafted and tested to capture certain market patterns or trading signals. Because these alphas are publicly available and widely used, they provide a strong and standardized benchmark for comparing with LLM-generated features.

To use these alphas more effectively, we also apply the Predictive Power Score (PPS) to measure how well each alpha predicts the stock closing price. 

Table~\ref{tab:top5 human-defined alphas} presents the top five human-defined formulaic alphas for each company, selected from the \textit{"101 Formulaic Alphas"}. The corresponding mathematical expressions for these alphas are provided in Table~\ref{tab: human-defined alpha_expressions}. As shown in the tables, none of the selected alphas incorporate sentiment information. Instead, they primarily rely on traditional market signals such as mean reversion (e.g., Alpha 5, 24, 57), momentum (e.g., Alpha 9, 10, 28), volatility and correlation (e.g., Alpha 18, 82), and relative strength (e.g., Alpha 42). Additionally, the structural diversity of these alphas is relatively limited.

Table~\ref{tab:combined_mse_comparison} presents the stock price prediction results of various models using three different types of alphas: LLM-generated, Featuretools-generated, and human-defined. Obviously, the prediction performance deteriorates significantly when using human-defined alphas as input with other conditions the same, compared to using LLM-generated alphas and featuretools-generated alphas. This performance gap underscores the advantage of leveraging LLMs to generate formulaic alphas because LLM-generated alphas can dynamically adapt to the latest stock movements by incorporating up-to-date information, whereas manually crafted alphas are typically static and prone to alpha decay, gradually losing their predictive power over time.

\begin{table*}[htbp]
\centering
\caption{Top Five Human-Defined Formulaic Alphas (Based on Predictive Power Score)}
\begin{tabular}{lccccc}
\toprule
\textbf{Rank} & \textbf{Apple} & \textbf{HSBC} & \textbf{Pepsi} & \textbf{Tencent} & \textbf{Toyota} \\
\midrule
1 & Alpha\#18 & Alpha\#18 & Alpha\#18 & Alpha\#18 & Alpha\#18 \\
2 & Alpha\#5 & Alpha\#42 & Alpha\#42  & Alpha\#42 & Alpha\#24 \\
3 & Alpha\#42 & Alpha\#57 & Alpha\#24 & Alpha\#24  & Alpha\#9  \\
4 & Alpha\#24  & Alpha\#28 & Alpha\#83 & Alpha\#83 & Alpha\#10 \\
5 & Alpha\#9 & Alpha\#33 & Alpha\#9 & Alpha\#9 & Alpha\#42  \\
\bottomrule
\end{tabular}
\label{tab:top5 human-defined alphas}
\end{table*}

\begin{table*}[h]
\renewcommand{\arraystretch}{1.4}
\small
\caption{Alpha Expressions Corresponding to Table~\ref{tab:top5 human-defined alphas}}
\label{tab: human-defined alpha_expressions}
\begin{tabularx}{\textwidth}{>{\bfseries}l>{\raggedright\arraybackslash}X}
\toprule
\textbf{Alpha ID} & \textbf{Expression} \\
\midrule
Alpha\#5 & $\text{rank}((\text{open} - (\text{sum}(\text{vwap}, 10) / 10))) \times (-1 \times \text{abs}(\text{rank}((\text{close} - \text{vwap}))))$ \\

Alpha\#9 & $(0 < \text{ts\_min}(\text{delta}(\text{close}, 1), 5)) \ ? \ \text{delta}(\text{close}, 1) : ((\text{ts\_max}(\text{delta}(\text{close}, 1), 5) < 0) ? \text{delta}(\text{close}, 1) : (-1 \times \text{delta}(\text{close}, 1)))$ \\

Alpha\#10 & $\text{rank}((0 < \text{ts\_min}(\text{delta}(\text{close}, 1), 4)) \ ? \ \text{delta}(\text{close}, 1) : ((\text{ts\_max}(\text{delta}(\text{close}, 1), 4) < 0) ? \text{delta}(\text{close}, 1) : (-1 \times \text{delta}(\text{close}, 1))))$ \\

Alpha\#18 & $-1 \times \text{rank}((\text{stddev}(\text{abs}(\text{close} - \text{open}), 5) + (\text{close} - \text{open})) + \text{correlation}(\text{close}, \text{open}, 10))$ \\

Alpha\#24 & $(((\text{delta}((\text{sum}(\text{close}, 100)/100), 100) / \text{delay}(\text{close}, 100)) < 0.05) \parallel ((\text{delta}((\text{sum}(\text{close}, 100)/100), 100) / \text{delay}(\text{close}, 100)) == 0.05)) \ ? \ (-1 \times (\text{close} - \text{ts\_min}(\text{close}, 100))) : (-1 \times \text{delta}(\text{close}, 3))$ \\

Alpha\#28 & $\text{scale}(\text{correlation}(\text{adv20}, \text{low}, 5) + ((\text{high} + \text{low}) / 2) - \text{close})$ \\

Alpha\#33 & $\text{rank}(-1 \times (1 - (\text{open} / \text{close}))^1)$ \\

Alpha\#42 & $\text{rank}(\text{vwap} - \text{close}) / \text{rank}(\text{vwap} + \text{close})$ \\

Alpha\#57 & $0 - \left(1 \times \frac{\text{close} - \text{vwap}}{\text{decay\_linear}(\text{rank}(\text{ts\_argmax}(\text{close}, 30)), 2)}\right)$ \\

Alpha\#82 & 
\parbox[t]{\linewidth}{
\begin{align*}
&\min\Big( \text{rank}(\text{decay\_linear}(\text{delta}(\text{open}, 1.46063), 14.8717)), \\
&\quad\text{Ts\_Rank}(\text{decay\_linear}(\text{correlation}(\text{IndNeutralize}(\text{volume}, \text{IndClass.sector}), \\
&\quad\quad((\text{open} \times 0.634196) + (\text{open} \times (1 - 0.634196))), 17.4842), 6.92131), 13.4283)\times -1\Big) 
\end{align*}
} \\
\bottomrule
\end{tabularx}
\end{table*}

\begin{table*}[htbp]
\centering
\caption{MSE Comparison of Models for Stock Price Prediction with Different Alpha Sources}
\label{tab:combined_mse_comparison}
\begin{tabular}{@{}llccccc@{}}
\toprule
\textbf{Alpha Source} & \textbf{Model} & \textbf{Apple} & \textbf{HSBC} & \textbf{Pepsi} & \textbf{Tencent} & \textbf{Toyota} \\
\midrule
\multirow{5}{*}{LLM-Generated} 
    & Transformer    & 0.0003 & 0.0004 & 0.0004 & 0.0002 & 0.0001 \\
    & Informer       & 0.0003 & 0.0007 & 0.0006 & 0.0004 & 0.0002 \\
    & LSTM           & 0.0030 & 0.0009 & 0.0018 & 0.0022 & 0.0015 \\
    & TCN            & 0.0018 & 0.0006 & 0.0028 & 0.0077 & 0.0046 \\
    & SVR            & 0.0048 & 0.0024 & 0.0046 & 0.0052 & 0.0048 \\
    & Random Forest  & 0.0019 & 0.0004 & 0.0014 & 0.0011 & 0.0005 \\
    & XGBoost  & 0.0019 & 0.0005 & 0.0014 & 0.0011 & 0.0005 \\
\midrule
\multirow{5}{*}{Featuretools-Generated} 
    & Transformer    & 0.0013 & 0.0019 & 0.0014 & 0.0015 & 0.0012 \\
    & Informer       & 0.0006 & 0.0050 & 0.0013 & 0.0011 &  0.0009 \\
    & LSTM           & 0.0052 & 0.0016 & 0.0026 & 0.0040 & 0.0025 \\
    & TCN            & 0.0048 & 0.0018 & 0.0031 & 0.0042 & 0.0024 \\
    & SVR            & 0.0077 & 0.0035 & 0.0030 & 0.0061 & 0.0058 \\
    & Random Forest  & 0.0035 & 0.0016 & 0.0032 & 0.0030 & 0.0025 \\
    & XGBoost  & 0.0022 & 0.0091 & 0.0087 & 0.0016 & 0.0019 \\
\midrule
\multirow{5}{*}{Human-Defined} 
    & Transformer    & 0.1923 & 0.0975 & 0.0862 & 0.1287 & 0.1142 \\
    & Informer       & 0.2924 & 0.1311 & 0.2471 & 0.0481 & 0.1424  \\

    & LSTM           & 0.2402 & 0.1359 & 0.2203 & 0.2224 & 0.1053 \\
    & TCN            & 0.1907 & 0.1285 & 0.2089 & 0.2049 & 0.1102 \\
    & SVR            & 0.2033 & 0.1555 & 0.4289 & 0.4315 & 0.3159 \\
    & Random Forest  & 0.1864 & 0.1308 & 0.1978 & 0.1964 & 0.0969 \\
    & XGBoost  & 0.2082 & 0.1428 & 0.2172 & 0.1481 & 0.1170 \\
\bottomrule
\end{tabular}
\end{table*}

\subsection{Further Evaluation of LLM-Generated Alphas}

In addition to traditional error-based evaluation metrics such as MSE, the \textbf{Information Coefficient (IC)} is commonly used in the financial domain to assess the predictive quality of return forecasts~\cite{zhang2020informationcoefficientperformancemeasure}, often referred to as alpha signals. IC quantifies the correlation between predicted returns and actual future returns, offering insight into the directional accuracy of an alpha’s forecasts. 

The value of IC ranges from \(-1\) to \(+1\). An IC of \(+1\) indicates a perfect positive linear relationship between an alpha signal and the realized returns, meaning that the alpha ranks the assets in exactly the same order as their actual performance. An IC of \(-1\) implies a perfect inverse relationship, where the alpha ranks the assets in the opposite order of their realized returns. An IC of \(0\) suggests that there is no linear correlation, indicating that alpha does not provide meaningful predictive information.

The IC is typically calculated using the Pearson correlation coefficient, and is given by:

\begin{equation}
\text{IC} = \frac{\text{Cov}(\hat{r}_t, r_t)}{\sigma_{\hat{r}_t} \sigma_{r_t}}
\end{equation}

\noindent where \( \hat{r}_t \) denotes the predicted return at time \( t \), \( r_t \) represents the actual realized return, and \( \sigma \) denotes the standard deviation. 

Due to its ability to reflect both the strength and direction of predictive signals, IC is widely adopted for evaluating the performance of quantitative models in finance.

\begin{table*}[htbp]
\centering
\caption{Information Coefficient (IC) of Five Alphas (Company-Specific)}
\label{tab:ic_results}
\begin{tabular}{lccccc}
\toprule
\textbf{Alpha} & \textbf{Apple} & \textbf{HSBC} & \textbf{Pepsi} & \textbf{Tencent} & \textbf{Toyota} \\
\midrule
Alpha 1 & $-0.0321$ & 0.0345 & $-0.0419$ & $-0.0105$ & $-0.0267$ \\
Alpha 2 & $-0.0114$ & 0.0307 & $-0.0226$ & 0.0193 & 0.0169 \\
Alpha 3 & $-0.0152$ & 0.0330 & 0.0009 & 0.0322 & 0.0092  \\
Alpha 4 & 0.0210 & 0.0124 & $-0.0018$ & $-0.0128$ & 0.0142 \\
Alpha 5 & $-0.0028$ & 0.0191 & 0.0551 & $-0.0149$ & 0.0140 \\
\bottomrule
\end{tabular}
\end{table*}

The IC values for five alphas of five different companies (Table~\ref{tab:ic_results}) show a mix of both positive and negative correlations. Importantly, each company has its own unique set of five alphas, meaning that Alpha 1 for Apple is not the same as Alpha 1 for Pepsi, despite sharing the same label structure.

Most of the alphas tend to produce relatively low IC values, indicating that their predictive power of stock return is weak or inconsistent. 

Overall, the LLM-generated alphas demonstrate weak predictive power for stock returns but exhibit strong predictive ability for stock prices, which is demonstrated in Table~\ref{tab:mse_comparison}. This discrepancy may stem from the fact that stock price prediction is more straightforward for LLMs, given their ability to process large volumes of relevant data, while return prediction involves additional complexity, requiring a deeper analysis of the relative changes in stock value over time.

\section{Conclusion}\label{sec5}

This study proposes a novel framework that integrates LLM-generated formulaic features with a Transformer model to forecast stock movement. The formulaic alphas are derived from basic inputs, such as stock features, technical indicators, and sentiment scores. For sentiment data, both target companies and their related companies are considered. Related companies are initially extracted using NER, based on their frequency of occurrence in news articles about the target companies. The most relevant ones, those strongly correlated with stock movement, are further selected by the LLM and incorporated into the formulaic alphas. For comparison of performance, models including Informer, LSTM, TCN, SVR, Random Forest, and XGBoost are used. The results highlight the predictive strength of the LLM-generated features.

This study has several limitations. First, the analysis is conducted on only five companies and within a single evaluation period. Future studies may consider a larger set of stocks and multiple time periods to improve generalizability. In addition, more recent and advanced time-series architectures such as N-BEATS can be included as baseline models for comparison. Besides, it can be explored how the number of LLM-generated formulaic alphas will affect prediction performance. Different prompt designs to generate formulaic alphas and different LLMs can be used for comparison.

\bibliographystyle{unsrt}  
\bibliography{references}

\end{document}